\documentclass[aip,reprint,superscriptaddress,amsmath,amssymb,twocolumn,preprintnumbers]{revtex4-2}

\usepackage{graphics,graphicx}

\begin{document}
\preprint{Physics of Fluids}

\title{Momentum flux fluctuations in wall turbulence formulated along the distance from the wall}

\author{Hideaki Mouri}
\affiliation{Meteorological Research Institute, Nagamine, Tsukuba 305-0052, Japan}
\author{Junshi Ito}
\affiliation{Meteorological Research Institute, Nagamine, Tsukuba 305-0052, Japan}
\affiliation{Graduate School of Science, Tohoku University, Aoba, Sendai 980-8578, Japan}


%

\begin{abstract}

Wall turbulence has a sublayer where the mean wall-normal flux of the streamwise momentum is constant.~Via the law of the wall, this mean flux is related to the wall-normal profile of the mean streamwise velocity. However, the momentum flux has large fluctuations, for which the corresponding law is yet unknown. To formulate such a law, we decompose fluctuations of the streamwise and the wall-normal velocities. These are smoothed to single out a component that would dominate the momentum flux fluctuations. It is dependent on the wall-normal distance. We relate this dependence to the wall-normal profile of the streamwise velocity variance. The resultant law is consistent with laboratory and field data across a wide range of that distance and applies readily to wall modeling of a numerical simulation.

\end{abstract}

\maketitle

\section{Introduction} \label{S1}
\vspace{-0.2mm} %

Wall turbulence transfers the momentum of the flow to the wall surface. The mean rate of this transfer, i.e., the mean momentum flux, is related to the mean streamwise velocity via the so-called law of the wall.\cite{my71} However, between fluctuations of the momentum flux and those of the streamwise velocity, such a law is yet unknown. Our understanding of the wall turbulence is thus incomplete. This incompleteness hinders, e.g., modeling of a numerical simulation. Accordingly, we explore that law on the basis of our previous studies of Refs.~\onlinecite{mmh19, im21, mi22}.

The turbulence is assumed to be stationary. Averages are those over time $t$. For example, $X(t)$ is averaged as $\langle X \rangle = \lim_{T \rightarrow +\infty} \int^{+T/2}_{-T/2} X(t)dt/T$.

\begin{figure}[bp]
\begin{center}
\resizebox{7.9cm}{!}{\includegraphics*[2.7cm,14.7cm][18.2cm,22.7cm]{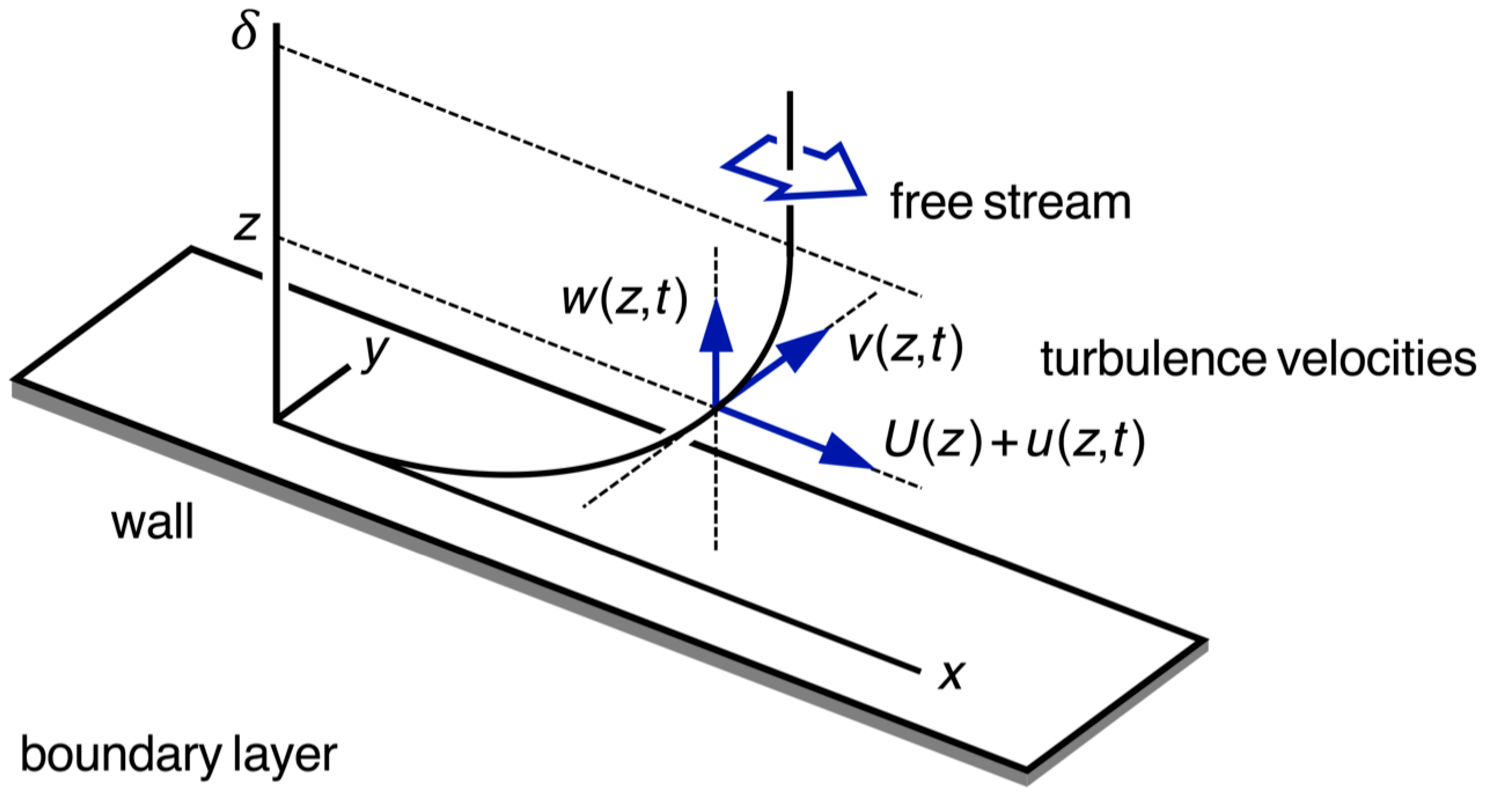}}
\caption{\label{f1} Schematic of a boundary layer as an example of wall turbulence. We indicate the $x$, $y$, and $z$ axes, the turbulence velocities $U(z)+u(z,t)$, $v(z,t)$, and $w(z,t)$, where $t$ is the time, and also the boundary layer thickness $\delta$. The fluctuations of the momentum flux $uw$ are described by our formulae of Eq.~(\ref{eq2}).}
\end{center}
\end{figure}

As illustrated in Fig.~\ref{f1}, we take the $x$--$y$ plane at the wall surface. The $x$ axis is in the mean flow direction. For a position $(x,y)$, while $U(z)$ denotes the mean velocity at a distance $z$ from that surface, $u(z,t)$, $v(z,t)$, and $w(z,t)$ respectively denote velocity fluctuations in the $x$, $y$, and $z$ directions. In addition, $\delta$ denotes the thickness of the turbulence region.

The surface of the wall is either smooth or rough. A rough surface is described by the aerodynamic roughness length $z_0$. To consider analogously a smooth surface, we parameterize it as $z_0 \varpropto \nu/u_{\ast}$.\cite{my71} Here $\nu$ is the kinematic viscosity and $u_{\ast}$ is the friction velocity. For turbulence at $z \gg z_0$, it is not essential whether the surface is smooth or rough.

We assume the incompressibility of the flow and study its momentum flux per unit mass density. The component of our main interest is $uw$, which corresponds to the transfer of the streamwise momentum in the wall-normal direction.

By taking the limit $\delta /z_0 \rightarrow +\infty$, there is attained a sublayer at $z_0 \ll z \ll \delta$ in which the mean momentum flux $\langle uw \rangle$ is constant along that distance $z$, i.e., $\langle uw \rangle = -u_{\ast}^2 < 0$.\cite{my71} Even if $\delta /z_0$ is finite, such a constant-flux sublayer remains at least as an approximation.

This sublayer obeys the law of the wall. While the friction velocity $u_{\ast}$ serves as a characteristic velocity, there is no constant in length. Then,\cite{my71}
\begin{subequations} \label{eq1}
\begin{equation} \label{eq1a}
\frac{z}{u_{\ast}} \frac{dU}{dz} = \frac{1}{\kappa} .
\end{equation}
For the von K\'arm\'an constant $\kappa$, we use a value $0.40$.\cite{mmhs13} An integration of Eq.~(\ref{eq1a}) yields the law of the wall as
\begin{equation} \label{eq1b}
\frac{U(z)}{u_{\ast}} = \frac{1}{\kappa} \ln \left( \frac{z}{z_0} \right)
\ \ \mbox{or} \ \
\langle uw \rangle = -u_{\ast}^2 = - \left[ \frac{\kappa U(z)}{\ln (z/z_0)} \right]^2 \! \! .
\end{equation}
\end{subequations}
The mean momentum flux $\langle uw \rangle$ is thus related to the profile of the mean velocity $U(z)$.

However, in the constant-flux sublayer, the momentum flux $uw$ fluctuates largely in time $t$.\cite{my71} It is not always inward as $uw < 0$ and is often outward as $uw > 0$. The law of these fluctuations is still uncertain but would be crucial to understanding the wall turbulence,\cite{my71, web72, wl72, wb77, amt00, nkpk07, dn11b, w16, dm21} especially about fluctuations of the streamwise velocity $U+u$. That law would also apply to wall modeling of a numerical simulation.\cite{d70, bmp05, kl12, lkbb16, ypm17, bp18, bl21, bogcn21}

We explore a law of fluctuations of the momentum flux $uw$ that would hold anywhere in the constant-flux sublayer. The result is
\begin{subequations} \label{eq2}
\begin{equation} \label{eq2a}
\overline{uw}_{\tau}(z,t) - \langle uw \rangle = -K(z) \left[ \overline{u^2}_{\tau}(z,t) - \langle u^2(z) \rangle \right] ,
\end{equation}
with smoothing over a timescale $\tau$ larger than a few tens of $z/U$,\cite{im21}
\begin{equation} \label{eq2b}
\overline{X}_{\tau}(z,t) = \frac{1}{\tau} \int^{+\tau/2}_{-\tau/2} X(z, t+t') dt' .
\end{equation}
Thus, $\overline{uw}_{\tau}$ fluctuates around the mean flux $\langle uw \rangle$ in proportion to the deviation of $\overline{u^2}_{\tau}$ from its average $\langle u^2 \rangle$. The proportionality factor $K$ depends on the distance $z$ as
\begin{equation} \label{eq2c}
K(z) = \kappa \frac{\langle uw \rangle / \langle u^2(z) \rangle}{\langle uw \rangle / \langle u^2(z_{\rm m}) \rangle} 
     = \kappa \frac{\langle u^2(z_{\rm m}) \rangle /u_{\ast}^2}{\langle u^2(z) \rangle / u_{\ast}^2}.
\end{equation}
\end{subequations}
Here $z_{\rm m}$ is the middle distance of that constant-flux sublayer, at which $K$ is equal to the von K\'arm\'an constant~$\kappa$. For the other distances $z$, we correct the factor $K$ by using $\langle u^2(z) \rangle / u_{\ast}^2$ and $\langle u^2(z_{\rm m}) \rangle / u_{\ast}^2 \simeq 4.5$.\cite{hvbs12, mmhs13, mkbm15, ofsbta17, smhffhs18, mmym20, ofwkt22} The mean flux $\langle uw \rangle = -u_{\ast}^2$ is to be obtained via Eq.~(\ref{eq1b}).

We derived Eq.~(\ref{eq2a}) in Ref.~\onlinecite{mi22} but with $K$ identical to the constant $\kappa$. This law was consistent with laboratory data of a boundary layer at around the middle distance $z_{\rm m}$ or for $z/\delta = 0.1$ to $0.2$. Elsewhere, however,~$K = \kappa$ would not hold so that we would over- or underestimate the fluctuations of $\overline{uw}_{\tau} - \langle uw \rangle$. By also using data for a much smaller value of $z/\delta = 0.002$ from our field observation of the atmospheric boundary layer, we now reconsider Eq.~(\ref{eq2a}) with $K$ in the form of Eq.~(\ref{eq2c}).

The formulation of Eq.~(\ref{eq2}) is described in Sec.~\ref{S2}. For completeness, we derive Eq.~(\ref{eq2a}) as well as Eq.~(\ref{eq2c}). The laboratory and field data used here are those of our previous studies, Refs.~\onlinecite{mmh19} and \onlinecite{mi22} (Sec.~\ref{S3}). They confirm Eq.~(\ref{eq2}) in Sec.~\ref{S4}. We interpret Eq.~(\ref{eq2}) on a phenomenological model of energy-containing eddies\cite{t76, pc82, mouri17, mm19, mmym20, hyz20} and discuss its application to wall modeling of a numerical simulation\cite{d70, bmp05, kl12, lkbb16, ypm17, bp18, bl21, bogcn21} (Sec.~\ref{S5}). Finally, in Sec.~\ref{S6}, we conclude with a remark on an extension of Eq.~(\ref{eq2}) to thermally stratified wall turbulence.\cite{my71, f06, m14, ypa20}

\begin{figure}[tbp]
\begin{center}
\resizebox{8.8cm}{!}{\includegraphics*[3.9cm,16.0cm][18.5cm,26.4cm]{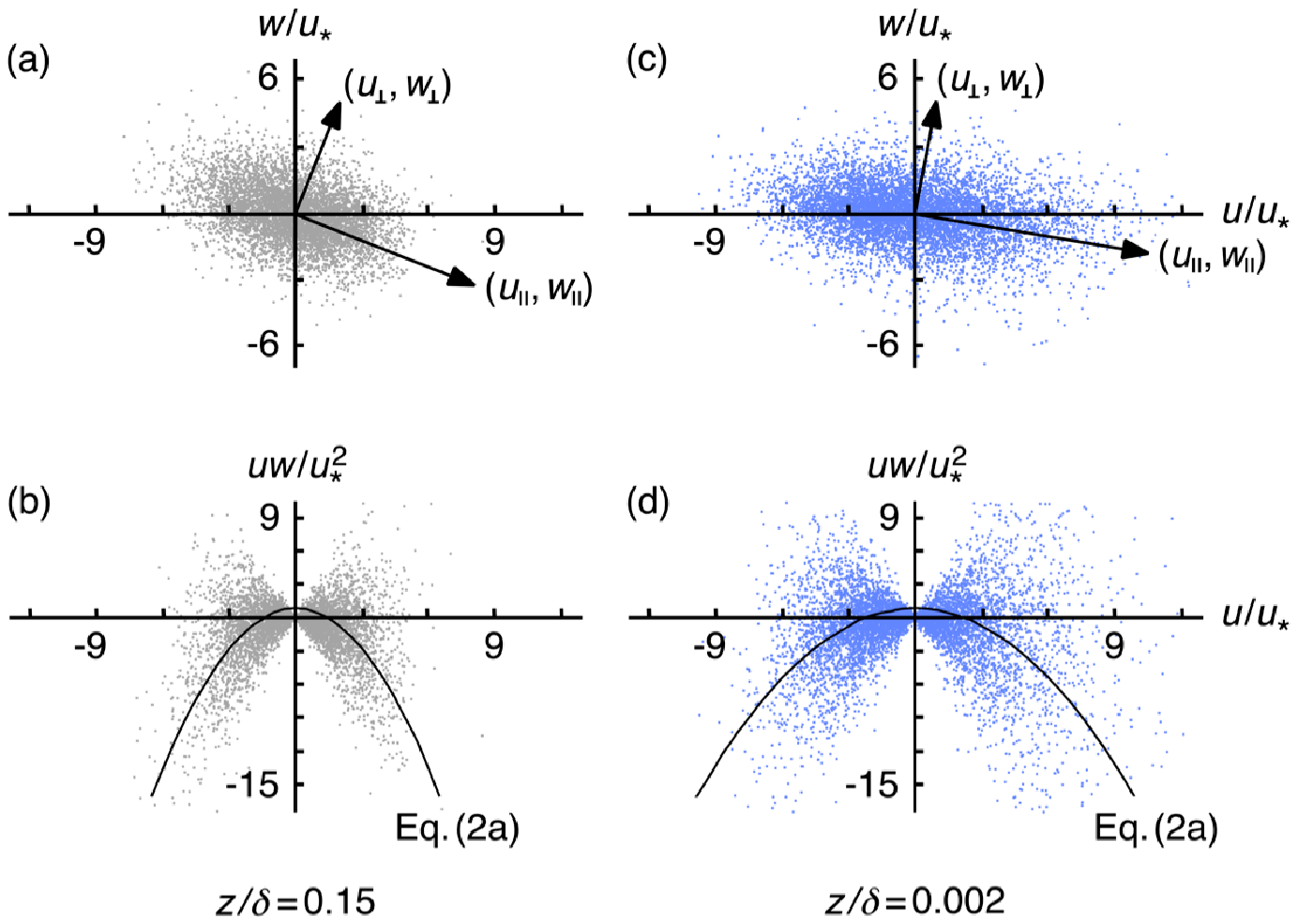}}
\caption{\label{f2} Scatter plots of $w/u_{\ast}$ (a) and $uw/u_{\ast}^2$ (b) against $u/u_{\ast}$ at $z/\delta = 0.15$ along with the vectors of Eq.~(\ref{eq3}) and the curve of Eq.~(\ref{eq2a}) for $K = 0.40$ and $\tau = 0$. Those at $z/\delta = 0.002$ for $K = 0.17$ are shown in (c) and (d).}
\end{center}
\end{figure} 

\section{Formulation} \label{S2}

Consider fluctuating velocities $u$ and $w$ at a given distance $z$ in the constant-flux sublayer. They scatter elliptically on the $u$--$w$ plane.\cite{im21, mi22, my71, wb77, w16} As observed in Figs.~\ref{f2}(a) and \ref{f2}(c), this ellipse has a major axis from the second quadrant $u<0$ and $w>0$ to the fourth quadrant $u>0$ and $w<0$, in accordance with the negative value of the mean momentum flux $\langle uw \rangle= -u_{\ast}^2 < 0$.

To describe such fluctuations, we decompose the vector $(u,w)$ at each time $t$ into orthogonal vectors $(u_{\shortparallel} , w_{\shortparallel})$ and $(u_{\perp} , w_{\perp})$.\! These two intend to align respectively with the major and the minor axes of the ellipse on the $u$--$w$ plane as in Figs.~\ref{f2}(a) and \ref{f2}(c),\cite{mi22}
\begin{subequations} \label{eq3}
\begin{equation} \label{eq3a}
u = u_{\shortparallel}+ u_{\perp} \ \ \mbox{and} \ \ w = w_{\shortparallel}+ w_{\perp} ,
\end{equation}
with
\begin{equation} \label{eq3b}
w_{\shortparallel} = -K(z) u_{\shortparallel} \ \ \mbox{and} \ \ w_{\perp} = u_{\perp}/K(z) .
\end{equation}
\end{subequations}
Here $K(z) > 0$ determines the directions of $(u_{\shortparallel} , w_{\shortparallel})$ and $(u_{\perp} , w_{\perp})$. The functional form of $K(z)$ is to hold at any distance $z$ in the constant-flux sublayer. Then, we single out the major component $(u_{\shortparallel} , w_{\shortparallel})$. By definition, it would dominate the average and the fluctuations of the momentum flux $uw$.

The major component $(u_{\shortparallel} , w_{\shortparallel})$ corresponds to an outward displacement ${\mit\Delta}z = w_{\shortparallel}{\mit\Delta}t > 0$ of a low-velocity fluid parcel $U+{\mit\Delta}U$ with ${\mit\Delta}U = u_{\shortparallel} < 0$ or an inward displacement ${\mit\Delta}z < 0$ of a high-velocity fluid parcel with ${\mit\Delta}U > 0$ over some timescale ${\mit\Delta}t$.\cite{web72, wl72, amt00, nkpk07, dn11b, w16, dm21} Being dominant in the mean momentum flux $\langle uw \rangle < 0$, such parcels would obey the mean velocity gradient $dU/dz = u_{\ast}/ \kappa z$ of Eq.~(\ref{eq1a}), i.e., ${\mit\Delta}U/{\mit\Delta}z =$ $-u_{\ast}/\kappa z$. If this is the case, since $z/u_{\ast}$ serves as the timescale ${\mit\Delta}t$, we have ${\mit\Delta}z / {\mit\Delta}t = -\kappa {\mit\Delta}U$, $w_{\shortparallel} = -\kappa u_{\shortparallel}$, and hence $K =$ $\kappa$.\cite{mi22}

However, in fact, ${\mit\Delta}z = w_{\shortparallel}{\mit\Delta}t \simeq z w_{\shortparallel} / u_{\ast}$ is not infinitesimal but could be comparable to the distance $z$. Around the middle $z \simeq z_{\rm m}$ of the constant-flux sublayer, ${\mit\Delta}z > 0$ and ${\mit\Delta}z < 0$ contribute almost equally to $dU/dz$ so that $K = \kappa$ holds as observed in Fig.~\ref{f2}(a). Elsewhere, we~need to correct the value of $K$.

This correction is on $\langle uw \rangle / \langle u^2(z) \rangle$, i.e., the slope of the regression line to data on the $u$--$w$ plane at each distance $z$. If the same fluctuations remain except for rescaling~of the $u$ and $w$ magnitudes, $K$ varies as $\varpropto \langle uw \rangle / \langle u^2(z) \rangle$. As a result,
\begin{equation} \label{eq4}
\frac{K(z)}{\kappa} 
=
\frac{\langle uw \rangle / \langle u^2(z) \rangle}{\langle uw \rangle / \langle u^2(z_{\rm m}) \rangle} 
=
\frac{\langle u^2(z_{\rm m}) \rangle /u_{\ast}^2}{\langle u^2(z) \rangle / u_{\ast}^2}
.
\end{equation}
We thereby obtain Eq.~(\ref{eq2c}). The middle distance $z_{\rm m}$ depends on the flow.\cite{hvbs12, mmhs13, mkbm15, ofsbta17, smhffhs18, mmym20, ofwkt22} From our laboratory data (Sec. \ref{S3a}), where the constant-flux sublayer lies from $z/\delta = 0.10$ to $0.20$, we adopt $z_{\rm m}/\delta = 0.15$ with $\langle u^2(z_{\rm m}) \rangle / u_{\ast}^2 = 4.67$. The vector $(u_{\shortparallel}, w_{\shortparallel})$ now always aligns with the major axis of the ellipse on the $u$--$w$ plane as in Figs.~\ref{f2}(a) and \ref{f2}(c).

Here $\langle uw \rangle / \langle u^2 \rangle$ is not equal to the inclination $w_{\shortparallel}/u_{\shortparallel} = -K$ of this ellipse, e.g., $\langle uw \rangle / \langle u^2 \rangle = -0.21$ but $w_{\shortparallel}/u_{\shortparallel} = -0.40$ in Fig.~\ref{f2}(a). Since $\langle uw \rangle / \langle u^2 \rangle$ might suffer from~an unidentified component that does not contribute to the wall-normal velocity $w$ (Sec.~\ref{S5a}),\cite{web72} we have used $dU/dz$ to obtain $\kappa$ as the value of $K(z_{\rm m})$.

On the other hand, because of $u_{\perp}w_{\perp} > 0$, the minor component $(u_{\perp} , w_{\perp})$ is regarded as random disturbances against the mean momentum flux $\langle uw \rangle < 0$. Since $(u_{\perp} , w_{\perp})$ does not fluctuate largely, its timescale is also not large. Actually, as per laboratory and field data,\cite{w16, web72, wl72, nkpk07, dm21} the fluctuations tend to have shorter durations in the first quadrant $u>0$ and $w>0$ and the third quadrant $u<0$ and $w<0$ than they do in the other quadrants.

We go on to derive Eq.~(\ref{eq2a}) on the basis of Ref.~\onlinecite{mi22}. By applying the smoothing of Eq. (\ref{eq2b}) to the decomposition of Eq.~(\ref{eq3a}),
\begin{subequations} \label{eq5}
\begin{equation} \label{eq5a}
\overline{uw}_{\tau}
=
  \overline{u_{\shortparallel}w_{\shortparallel}}\vert_{\tau}  
+ \overline{u_{\shortparallel}w_{\perp}         }\vert_{\tau}
+ \overline{u_{\perp}         w_{\shortparallel}}\vert_{\tau}  
+ \overline{u_{\perp}         w_{\perp}         }\vert_{\tau} . 
\end{equation}
Here $\overline{X}\vert_{\tau}$ is equivalent to $\overline{X}_{\tau}$. The same decomposition yields
\begin{equation} \label{eq5b}
\langle uw \rangle
=
  \langle u_{\shortparallel}w_{\shortparallel} \rangle  
+ \langle u_{\shortparallel}w_{\perp}          \rangle 
+ \langle u_{\perp}         w_{\shortparallel} \rangle  
+ \langle u_{\perp}         w_{\perp}          \rangle . 
\end{equation}
\end{subequations}
If fluctuations of the minor component $(u_{\perp}, w_{\perp})$ are independent of those of the major component $(u_{\shortparallel}, w_{\shortparallel})$ and also if the former fluctuations do have shorter durations, $\overline{u_{\shortparallel}w_{\perp}}\vert_{\tau}$, $\overline{u_{\perp}w_{\shortparallel}}\vert_{\tau}$, and $\overline{u_{\perp} w_{\perp}}\vert_{\tau}$ converge to their averages faster than $\overline{u_{\shortparallel}w_{\shortparallel}}\vert_{\tau}$. For an appropriate timescale $\tau$, we expect
\begin{subequations} \label{eq6}
\begin{equation} \label{eq6a}
\overline{u_{\shortparallel}w_{\perp}}\vert_{\tau} \rightarrow  \langle u_{\shortparallel}w_{\perp}          \rangle =0 , \ \ 
\overline{u_{\perp}w_{\shortparallel}}\vert_{\tau} \rightarrow  \langle u_{\perp}         w_{\shortparallel} \rangle =0 ,
\end{equation}
and
\begin{equation} \label{eq6b}
\overline{u_{\perp}w_{\perp}} \vert_{\tau} \rightarrow \langle u_{\perp}   w_{\perp} \rangle          
                                           =           \langle u_{\perp}^2           \rangle /K > 0 ,
\end{equation}
but yet
\begin{equation} \label{eq6c}
\overline{u_{\shortparallel}w_{\shortparallel}}\vert_{\tau} \nrightarrow \langle u_{\shortparallel}w_{\shortparallel} \rangle
                                                            =         -K \langle u_{\shortparallel}^2                 \rangle < 0 .
\end{equation}
\end{subequations}
From Eqs.~(\ref{eq5}) and (\ref{eq6}),
\begin{subequations} \label{eq7}
\begin{equation} \label{eq7a}
\overline{uw}_{\tau} - \langle uw \rangle \rightarrow \overline{u_{\shortparallel}w_{\shortparallel}}\vert_{\tau} -   \langle u_{\shortparallel}w_{\shortparallel} \rangle 
                                          =       -K  \overline{u_{\shortparallel}^2                }\vert_{\tau} + K \langle u_{\shortparallel}^2                 \rangle .
\end{equation}
Likewise,
\begin{equation} \label{eq7b}
\overline{u^2}_{\tau} - \langle u^2 \rangle \rightarrow \overline{u_{\shortparallel}^2}\vert_{\tau} - \langle u_{\shortparallel}^2 \rangle .
\end{equation}
\end{subequations}
By using Eq.~(\ref{eq7}), we relate $\overline{uw}_{\tau} - \langle uw \rangle$ with $\overline{u^2}_{\tau} - \langle u^2 \rangle$ to obtain Eq.~(\ref{eq2a}) as an asymptotic law.

Figures~\ref{f2}(b) and \ref{f2}(d) show the curves of Eq.~(\ref{eq2a}).~They are convex upward and remain negative unless $\langle u_{\perp}w_{\perp}\rangle > 0$ dominates over $\overline{u_{\shortparallel}w_{\shortparallel}}\vert_{\tau} < 0$ in $\overline{uw}_{\tau} = \overline{u_{\shortparallel}w_{\shortparallel}}\vert_{\tau} + \langle u_{\perp}w_{\perp}\rangle$. This is in contrast to the case of $uw$ (scattered points), where unsmoothed fluctuations of $(u_{\perp}, w_{\perp})$ often induce a significantly positive value. Nevertheless, Eq.~(\ref{eq2a}) does reproduce $\overline{uw}_{\tau}$ smoothed over an appropriate timescale $\tau$ (Sec.~\ref{S4}).

\section{Data} \label{S3}
\vspace{-1mm} %

Two datasets are analyzed here. One is for $z/\delta = 0.10$ to $0.20$ and from our experiment of a boundary layer in a wind tunnel (Ref.~\onlinecite{mi22}). The other is for $z/\delta = 0.002$ and from our field observation of the atmospheric boundary layer (Ref.~\onlinecite{mmh19}). While the details are described in Refs.~\onlinecite{mmh19} and \onlinecite{mi22}, we outline the data and compare their main characteristics in Table~\ref{t1} and Figs.~\ref{f3} and \ref{f4}.
\begin{table}[bp]
\vspace{-3mm} %
\begingroup
\squeezetable
\caption{\label{t1} Turbulence characteristics of our laboratory and field data. We also show $\pm 1 \sigma$ errors.}
\begin{ruledtabular}
\begin{tabular}{lccc}
Quantity                & Unit             &  Laboratory           & Field                           \\  \hline
Thickness $\delta$      & mm               &  $373.    \pm 2.$     &\!$863.  \pm 14.\,(\times 10^3)$ \\    
Friction velocity $u_*$ & mm\,s$^{-1}$     &  $542.    \pm 0.$     &  $247.  \pm  4.$                \\
Viscosity $\nu$         & mm$^2$\,s$^{-1}$ &  $ 15.3   \pm 0.0$    &  $ 13.8 \pm  0.0$               \\
Roughness $z_0$         & mm               &\!$  0.099 \pm 0.011$\!&  $ 20.9 \pm  0.7$               \\ 
\end{tabular}
\end{ruledtabular}
\endgroup
\end{table}

\vspace{-1mm} %
\subsection{Laboratory data} \label{S3a}
\vspace{-1mm} %

The laboratory experiment was in a wind tunnel of the Meteorological Research Institute. Its measurement section has the streamwise size ${\mit\Delta}x = 22$\,m, the spanwise size ${\mit\Delta}y = 3$\,m, and the floor-normal size ${\mit\Delta}z = 2$\,m. Upon the surface of the floor, we set spanwise rods of diameter $2$ mm with an interval ${\mit\Delta}x = 50$\,mm as the roughness.

The incoming flow velocity was $12$\,m\,s$^{-1}$. Over the distances $z \le 600$\,mm at ${\mit\Delta}x = 18$\,m downstream of the first spanwise rod, time series were obtained for the streamwise velocity $U+u$ and the surface-normal velocity $w$. We used a hot-wire anemometer with a crossed-wire sensor of size $1$\,mm. Especially at $z = 38$ to $75$\,mm in the constant-flux sublayer,\cite{mi22} where $U$ ranges from $7.87$ to $8.77$\,m\,s$^{-1}$, the sampling frequency was $40$\,kHz. The total length of each time series was $2.4 \times 10^8$.

We divide the time series into blocks of length $1 \times~10^7$, regard them as independent samples of the same turbulence, and calculate statistics over time $t$. Then, those of the turbulence, i.e., the parent population, are estimated along with the errors in a standard manner.\cite{br03} Our tables and figures show $\pm 1 \sigma$ errors, albeit not always evident. For the tables, note an additional uncertainty due to the finite number of the significant digits.

\begin{figure}[tbp]
\begin{center}
\rotatebox{270}{
\resizebox{4.7cm}{!}{\includegraphics*[2.9cm,2.8cm][15.8cm,26.4cm]{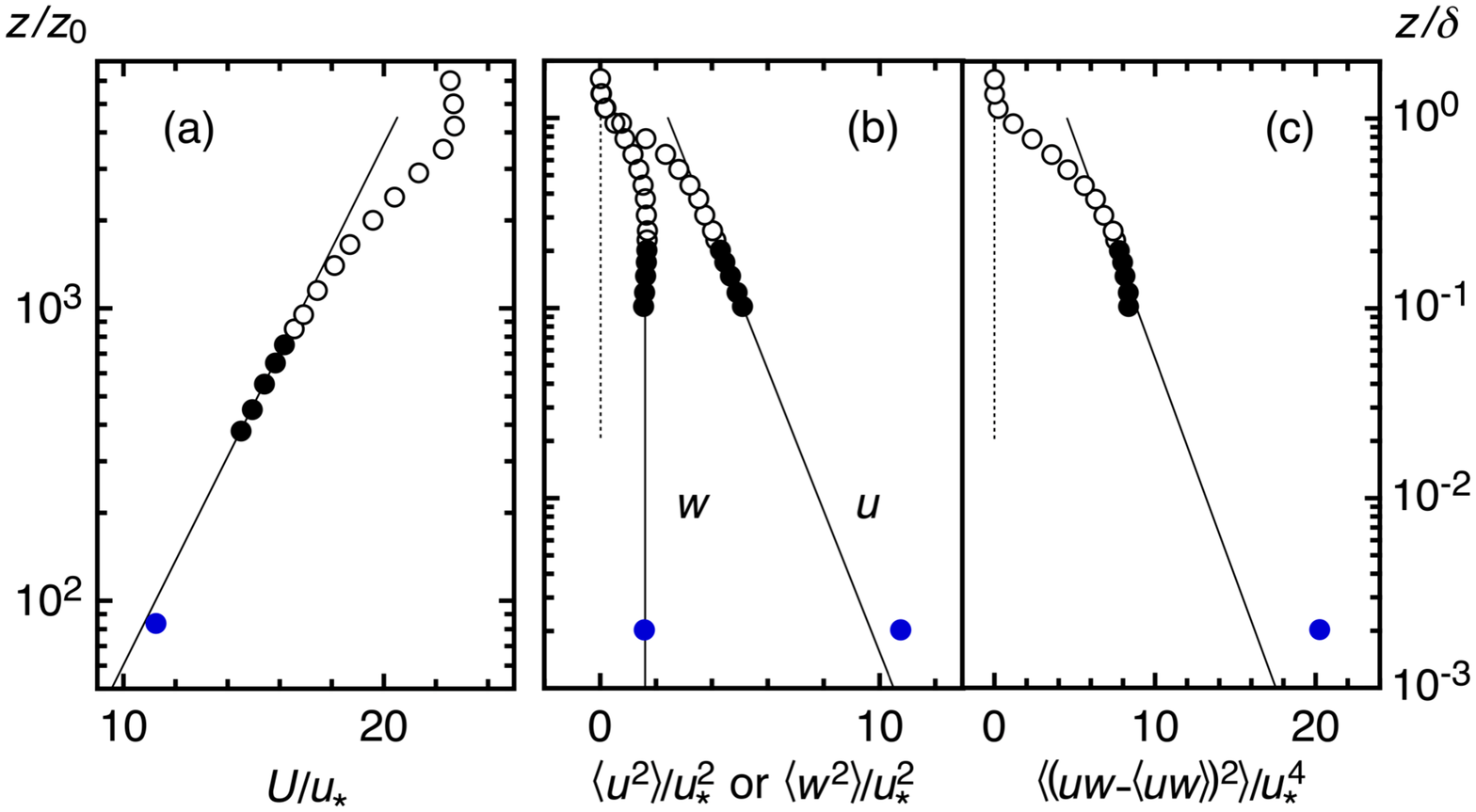}}
}
\caption{\label{f3} One-time statistics $U/u_{\ast}$ (a), $\langle u^2 \rangle/u_{\ast}^2$, $\langle w^2 \rangle/u_{\ast}^2$ (b), or $\langle (uw - \langle uw \rangle)^2 \rangle/u_{\ast}^4$ (c) against $z/z_0$ or $z /\delta$. The filled symbols lie in the constant-flux sublayer. Errors are smaller than the size of the symbols. The solid lines are regression fits of Eq. (\ref{eq1b}) or (\ref{eq8}) to the data at $z/z_0 = 380$ to $760$ or $z/\delta =$ $0.10$ to $0.20$.}
\vspace{1mm}
\resizebox{8.8cm}{!}{\includegraphics*[3.0cm,18.7cm][17.3cm,26.6cm]{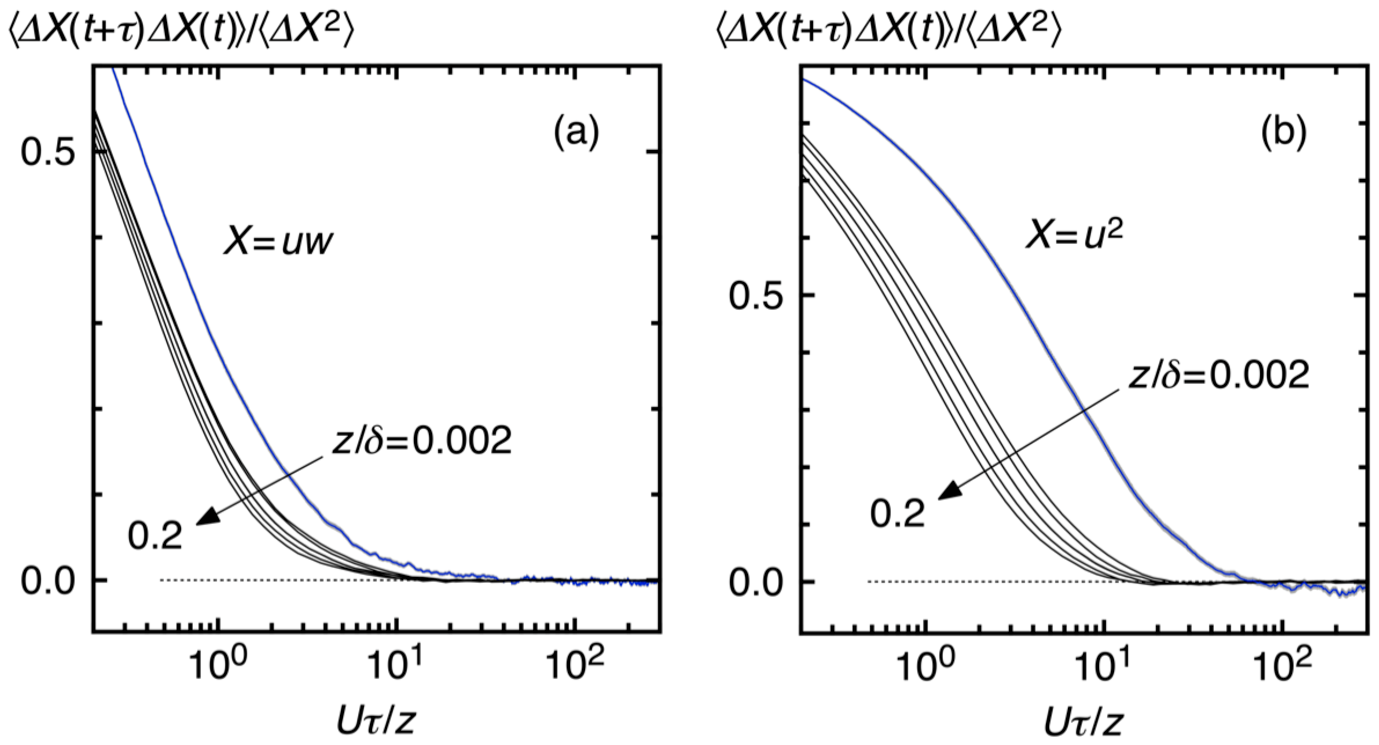}}
\caption{\label{f4} Two-time correlation $\langle {\mit\Delta}X(t+\tau){\mit\Delta}X(t) \rangle / \langle {\mit\Delta}X^2 \rangle$ of ${\mit\Delta}X = X - \langle X \rangle = uw - \langle uw \rangle$ (a) or $u^2 - \langle u^2 \rangle$ (b) against $U\tau/z$ at $z/\delta = 0.002$ and $0.10$ to $0.20$. For those at $z/\delta = 0.002$ and $0.15$, gray areas denote $\pm 1\sigma$ errors.}
\end{center}
\vspace{-3mm} %
\end{figure}

The friction velocity $u_{\ast}$ is estimated by averaging $\langle uw \rangle$ over the constant-flux sublayer. Its law of the wall of Eq. (\ref{eq1b}) holds as observed in Fig.~\ref{f3}(a). We also estimate the boundary layer thickness $\delta$ as the distance $z$ at which the mean streamwise velocity $U$ was equal to $99$\,\% of its maximum value.

\vspace{-1mm} %
\subsection{Field data} \label{S3b}
\vspace{-1mm} %

The observation was in the winter seasons from $2016$ to $2018$ at a flat grass field of the Meteorological Research Institute ($36.055^{\circ}$N, $140.123^{\circ}$E). During those seasons, the grass was dry and cut to height $\le 50$\,mm. The surface was almost the same over $\ge 300$\,m in the prevailing wind direction.

We measured the wind velocities $U+u$, $v$, and $w$ along with the air temperature at a distance $z = 1.75$\,m from the surface by using an ultrasonic anemometer. Its sensor size was about $100$\,mm. The sampling frequency was $10$ Hz.

We used several criteria to select data blocks such that the thermal stratification was neutral and neither the wind velocity nor the wind direction varied too much (see Ref.~\onlinecite{mmh19} for details). The same criteria are used here, except for $15$\,min instead of $30$\,min for the block length\cite{my75, llp16} and $3.0 \pm 1.0$\,m\,s$^{-1}$ instead of $3.0 \pm 1.5$\,m\,s$^{-1}$ for the mean velocity $U$. There are $145$ blocks of such data, in each of which we normalize the time $t$ by $z/U$ and the velocities $U$, $u$, $v$, and $w$ by the friction velocity $u_{\ast} = \langle -uw \rangle^{1/2}$ for the further analyses. Statistics and their errors are to be from these data blocks.

The boundary layer thickness is calculated in each data block as $\delta = 0.3 u_{\ast}/|f_C|$.\cite{cvh72, hs75, ssb13} Here $f_{\rm C} = 1.46 \times 10^{-4} \sin \varphi$ in units of rad\,s$^{-1}$ is the Coriolis parameter at the latitude $\varphi$ (see Ref.~\onlinecite{mmh19} for confirmation with simultaneous radiosonde observations). The average of the calculations is used for the estimate of $\delta$ in Table \ref{t1}.

\subsection{Characteristics of the data} \label{S3c}

Figures \ref{f3}(b) and \ref{f3}(c) show the variances $\langle u^2 \rangle$, $\langle w^2 \rangle$, and $\langle (uw - \langle uw \rangle )^2 \rangle$ at the distances of $z/\delta = 0.002$ in the field data and of $z/\delta = 0.1$ to $1.6$ in the laboratory data. As a value in the constant-flux sublayer (filled symbols), $z/\delta = 0.002$ is small and not yet attained in a laboratory. Thus, for that sublayer, we have a unique opportunity to study a wide range of $z/\delta$.

The constant-flux sublayer is known to exhibit scaling laws of the velocity variances $\langle u^2 \rangle$ and $\langle w^2 \rangle$,\cite{t76}
\begin{subequations} \label{eq8}
\begin{equation} \label{eq8a}
\frac{\langle u^2(z) \rangle}{u_{\ast}^2} = c_{uu} + d_{uu} \ln \left( \frac{\delta}{z} \right)
\ \ \mbox{and} \ \
\frac{\langle w^2(z) \rangle}{u_{\ast}^2} = c_{ww} .
\end{equation}
Our laboratory data at $z = 38$ to $75$\,mm or $z/\delta = 0.10$ to $0.20$ yield $c_{uu} = 2.43 \pm 0.10$, $d_{uu} = 1.17 \pm 0.03$, and $c_{ww} = 1.61 \pm 0.02$. They are typical of a boundary layer, $c_{uu} \simeq 2.0$--$2.5$, $d_{uu} \simeq 1.2$--$1.3$,\cite{mmhs13, smhffhs18, mmym20} and $c_{ww} \simeq 1.3$--$1.6$.\cite{amt00, mkbm15} Also for the variance of the momentum flux $uw$, as derived in Sec.~\ref{S5a},
\begin{equation} \label{eq8b}
\frac{\langle (uw(z) - \langle uw \rangle )^2 \rangle}{u_{\ast}^4} = c_{uwuw} +1 + c_{uu}c_{ww} + c_{ww}d_{uu} \ln \left( \frac{\delta}{z} \right) .
\end{equation}
\end{subequations}
Here $c_{uwuw}$ is $-0.37 \pm 0.20$. The other parameters are~the same as for the laws of $\langle u^2 \rangle$ and $\langle w^2 \rangle$ in Eq.~(\ref{eq8a}).

Extrapolations of those fits of Eq.~(\ref{eq8}) reproduce the variances of our field data at $z/\delta = 0.002$ (solid lines in Fig.~\ref{f3}). There are slight discrepancies because the constant-flux sublayer in a wind tunnel is not identical to that in the atmosphere. The latter has, e.g., a diurnal cycle.\cite{my75, llp16} With a decrease in $z/\delta$, while $\langle w^2 \rangle$ remains the same, $\langle u^2 \rangle$ and $\langle (uw - \langle uw \rangle )^2 \rangle$ become large, in accordance with the behaviors of the instantaneous data on the $u$--$w$ and the $u$--$uw$ planes of Fig.~\ref{f2}.

Figure \ref{f4} shows two-time correlations \!$\langle {\mit\Delta}X(t+\tau){\mit\Delta}X(t) \rangle$ normalized by the variances $\langle {\mit\Delta}X^2 \rangle$ for ${\mit\Delta}X = X - \langle X \rangle = uw - \langle uw \rangle$ and $u^2 - \langle u^2 \rangle$ at $z = 38$, $45$, $55$, $65$, and $75$\,mm or $z/\delta = 0.10$, $0.12$, $0.15$, $0.17$, and $0.20$ in the wind tunnel and at $z = 1.75$\,m or $z/\delta = 0.002$ in the atmosphere. The timescale $\tau$ is limited to a range from $0.2z/U$ to $300z/U$ by the sampling frequency, block length, and mean streamwise velocity of our field data.

The two-time correlations have larger extents if $z/\delta$ is smaller. While the correlations in the wind tunnel always disappear at a timescale $\tau \simeq 20z/U$ (see Appendix), that of $uw$ for $z/\delta = 0.002$ in the atmosphere is observed up to $\tau \simeq 30U/z$. Moreover, because of the diurnal cycle~of the atmosphere,\cite{my75, llp16} its $u^2$ correlation is observed at any large timescale $\tau$. These field data would not attain exact convergence of their statistics, although we are to ignore such a weak correlation.

\begin{figure}[tbp]
\begin{center}
\resizebox{8.8cm}{!}{\includegraphics*[2.9cm,3.9cm][17.2cm,26.6cm]{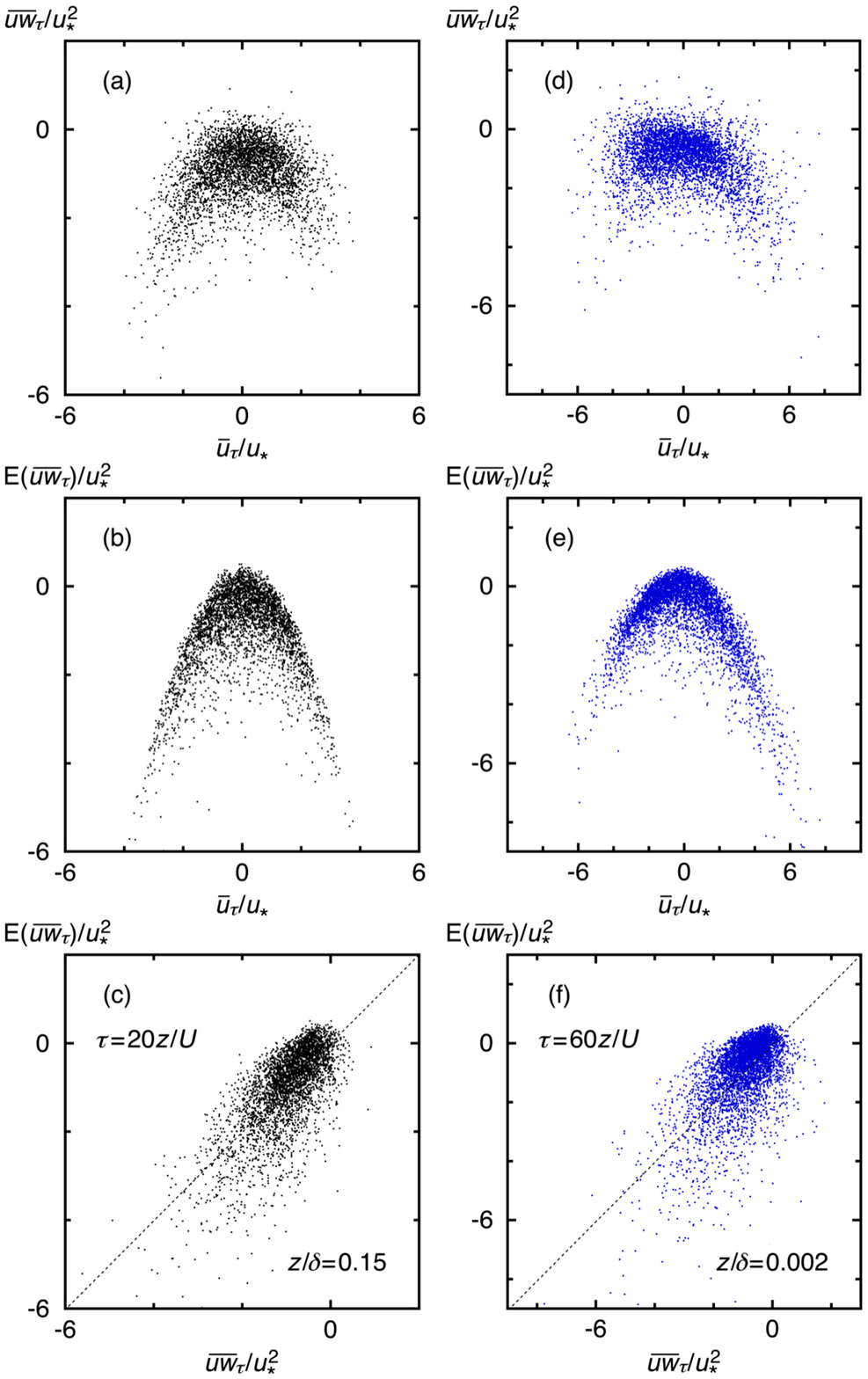}}
\caption{\label{f5} Scatter plots among $\overline{uw}_{\tau}/u_{\ast}^2$, ${\rm E}(\overline{uw}_{\tau})/u_{\ast}^2$, and $\overline{u}_{\tau}/u_{\ast}$ for $\tau = 20z/U$ at $z/\delta = 0.15$ (a--c) or for $\tau = 60z/U$ at $z/\delta =$ $0.002$ (d--f). Here ${\rm E}(\overline{uw}_{\tau})$ is an estimate of $\overline{uw}_{\tau}$ via~Eq. (\ref{eq2}).}
\end{center}
\end{figure} 

\section{Results} \label{S4}

To confirm the formulae of Eq.~(\ref{eq2}), we analyze our data of the constant-flux sublayer at $z/\delta = 0.10$ to $0.20$ in a wind tunnel and at $z/\delta = 0.002$ in the atmosphere (Sec. \ref{S3}). Their values of $\langle uw \rangle$ and $\langle u^2 \rangle$ determine the factor $K(z)$ via Eq.~(\ref{eq2c}). Along with $\overline{u^2}_{\tau}$ from Eq.~(\ref{eq2b}), those are used in Eq.~(\ref{eq2a}) to estimate the smoothed momentum flux at each time $t$ as ${\rm E}(\overline{uw}_{\tau})$. We compare this estimate with the actual value $\overline{uw}_{\tau}$.

Figure \ref{f5} shows scatter plots among $\overline{uw}_{\tau}$, ${\rm E}(\overline{uw}_{\tau})$, and the smoothed fluctuating streamwise velocity $\overline{u}_{\tau}$ at $z/\delta = 0.15$ and $0.002$. The smoothing timescale $\tau$ is $20z/U$ or $60z/U$, at which the two-time correlations of $uw$ and $u^2$ have become weak, if any, in Fig.~\ref{f4}.

Upon plots of $\overline{uw}_{\tau}$ against $\overline{u}_{\tau}$ in Figs.~\ref{f5}(a) and \ref{f5}(d), we observe a convex distribution such that $\overline{uw}_{\tau}$ tends to be positive when $\overline{u}_{\tau}$ is close to $0$ or equivalently $U+\overline{u}_{\tau}$ is close to $U$.\cite{im21} When $U+\overline{u}_{\tau}$ deviates from $U$ largely, $\overline{uw}_{\tau}$ is negative. These are respectively due to $\langle u_{\perp}w_{\perp}\rangle > 0$ and $\overline{u_{\shortparallel}w_{\shortparallel}}\vert_{\tau} < 0$ in Eqs.~(\ref{eq5})--(\ref{eq7}) as mentioned for the behavior of Eq.~(\ref{eq2a}) at $\tau = 0$ in Figs.~\ref{f2}(b) and \ref{f2}(d)

The same holds for ${\rm E}(\overline{uw}_{\tau})$ in Figs.~\ref{f5}(b) and \ref{f5}(e). Furthermore, upon plots of ${\rm E}(\overline{uw}_{\tau})$ against $\overline{uw}_{\tau}$ in Figs.~\ref{f5}(c) and \ref{f5}(f), the data points lie along a dotted line of $\overline{uw}_{\tau} =$ ${\rm E}(\overline{uw}_{\tau})$. Thus, ${\rm E}(\overline{uw}_{\tau})$ correlates with $\overline{uw}_{\tau}$ via the major component $(u_{\shortparallel} , w_{\shortparallel})$. The scatter of the data from that line is attributable to remaining fluctuations of the minor component $(u_{\perp}, w_{\perp})$.

These results in Fig.~\ref{f5} confirm the formulae of Eq.~(\ref{eq2}). As a further confirmation, we analyze statistics of the actual value $\overline{uw}_{\tau}$ and the estimate ${\rm E}(\overline{uw}_{\tau})$ of the smoothed momentum flux at $z/\delta = 0.002$, $0.10$, $0.12$, $0.15$, $0.17$, and $0.20$ against the smoothing timescale $\tau$.

\begin{figure}[bp]
\begin{center}
\resizebox{8.8cm}{!}{\includegraphics*[3.0cm,18.7cm][17.3cm,26.6cm]{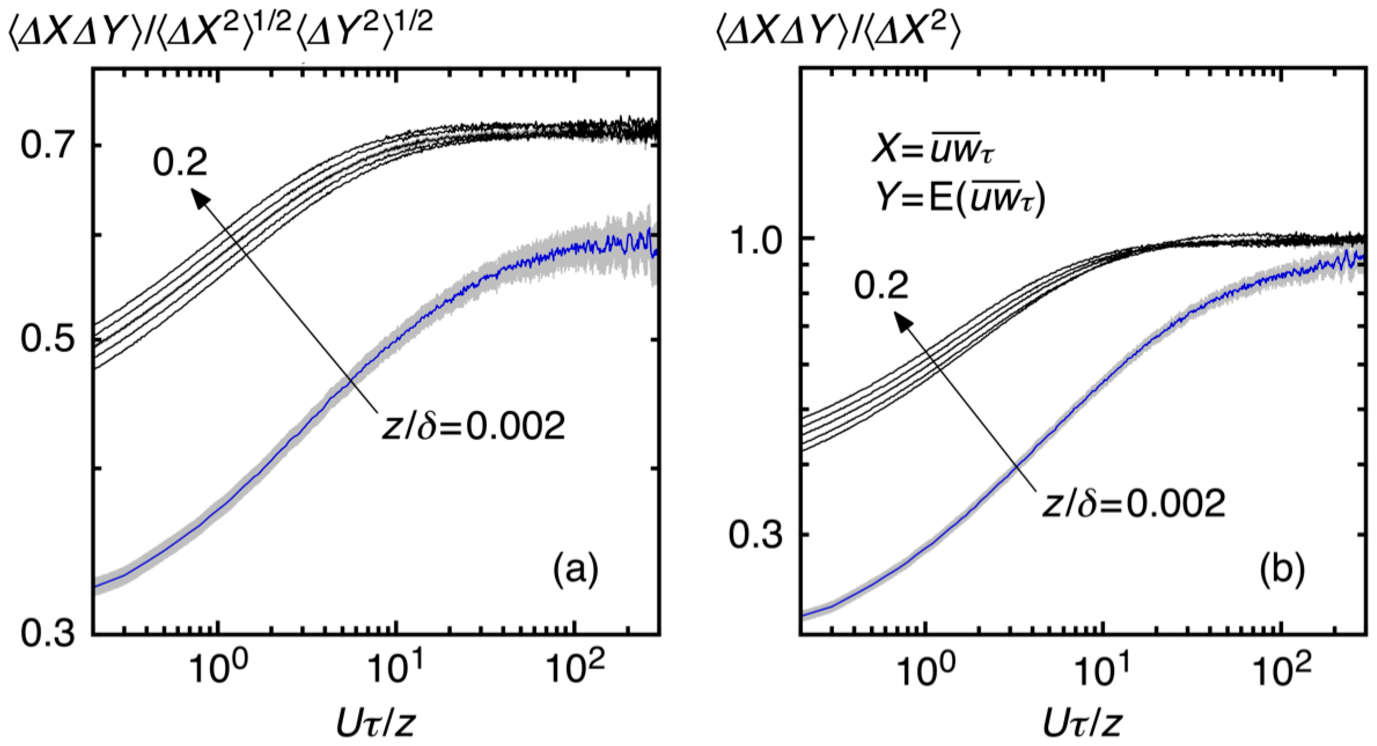}}
\caption{\label{f6} Coefficient $\langle {\mit\Delta}X{\mit\Delta}Y \rangle / \langle {\mit\Delta}X^2 \rangle^{1/2} \langle {\mit\Delta}Y^2 \rangle^{1/2}$ (a) and slope $\langle {\mit\Delta}X{\mit\Delta}Y \rangle / \langle {\mit\Delta}X^2 \rangle$ (b) of the correlation between ${\mit\Delta}X = X - \langle X \rangle = \overline{uw}_{\tau} - \langle \overline{uw}_{\tau} \rangle$ and ${\mit\Delta}Y = Y - \langle Y \rangle = {\rm E}(\overline{uw}_{\tau}) - \langle {\rm E}(\overline{uw}_{\tau}) \rangle$ against $U\tau/z$ at $z/\delta = 0.002$ and $0.10$ to $0.20$.~Here ${\rm E}(\overline{uw}_{\tau})$ is an estimate of $\overline{uw}_{\tau}$ via Eq.~(\ref{eq2}). For those at~$z/\delta = 0.002$ and $0.15$, gray areas denote $\pm 1\sigma$ errors.}
\end{center}
\end{figure} 

Figure~\ref{f6}(a) shows the correlation coefficient between $\overline{uw}_{\tau}$ and ${\rm E}(\overline{uw}_{\tau})$. With the timescale $\tau$, fluctuations of the minor component $(u_{\perp}, w_{\perp})$ are smoothed away. That coefficient becomes larger and then converges to a constant at $\tau \simeq 20z/U$ for $z/\delta = 0.20$ to $60z/U$ for $z/\delta =$ $0.002$. There, ${\rm E}(\overline{uw}_{\tau})$ is certainly related to $\overline{uw}_{\tau}$.

The correlation coefficient is yet less than unity, $\le 0.7$. Unlike our expectation in Sec.~\ref{S2}, some fluctuations of the minor component $(u_{\perp}, w_{\perp})$ have too large timescales to be smoothed away (Sec.~\ref{S5a}). Their fraction appears to increase with a decrease in $z/\delta$. Actually, from $z/\delta =$ $0.20$ to $0.10$ and from $0.10$ to $0.002$, the correlation coefficient is smaller and converges at a larger timescale $\tau$.

So long as $\overline{uw}_{\tau}$ is estimated from $U+u$ at the same distance $z$, since the available information is incomplete, the result is not exact. Previously, in Ref.~\onlinecite{im21}, we estimated $\overline{uw}_{\tau}$ from $U+u$ via machine learning of laboratory data. The correlation coefficient between $\overline{uw}_{\tau}$ and ${\rm E}(\overline{uw}_{\tau})$ was again $0.7$ around the middle $z \simeq z_{\rm m}$ of the constant-flux sublayer. Thus, Eq.~(\ref{eq2}) has attained the highest possible accuracy for such an estimation of the momentum flux $\overline{uw}_{\tau}$.

Figure~\ref{f6}(b) shows the correlation slope of ${\rm E}(\overline{uw}_{\tau})$ on $\overline{uw}_{\tau}$. The exact convergence is not observed for the field data at $z/\delta = 0.002$, owing to the aforementioned diurnal cycle of the atmosphere. Still, with the timescale $\tau$, each slope tends to unity, i.e., $\overline{uw}_{\tau} = {\rm E}(\overline{uw}_{\tau})$. That slope is $\varpropto K(z)$ as in Eq.~(\ref{eq2a}). Here $K(z) \varpropto u_{\ast}^2 / \langle u^2 (z) \rangle$ in Eq.~(\ref{eq2c}) varies from $0.17$ at $z/\delta = 0.002$ to $0.37$ at $z/\delta = 0.10$. From $z/\delta = 0.10$ to $0.20$, it varies from $0.37$ to $0.43$ (see Fig.~\ref{f3} for $\langle u^2 \rangle / u_{\ast}^2$). If $K(z)$ were constant at $\kappa = 0.40$,\cite{mi22} the slope would not necessarily tend to unity. We thereby confirm $K(z)$ in the form of Eq.~(\ref{eq2c}).

\begin{figure}[tbp]
\begin{center}
\resizebox{8.8cm}{!}{\includegraphics*[3.0cm,18.7cm][17.3cm,26.6cm]{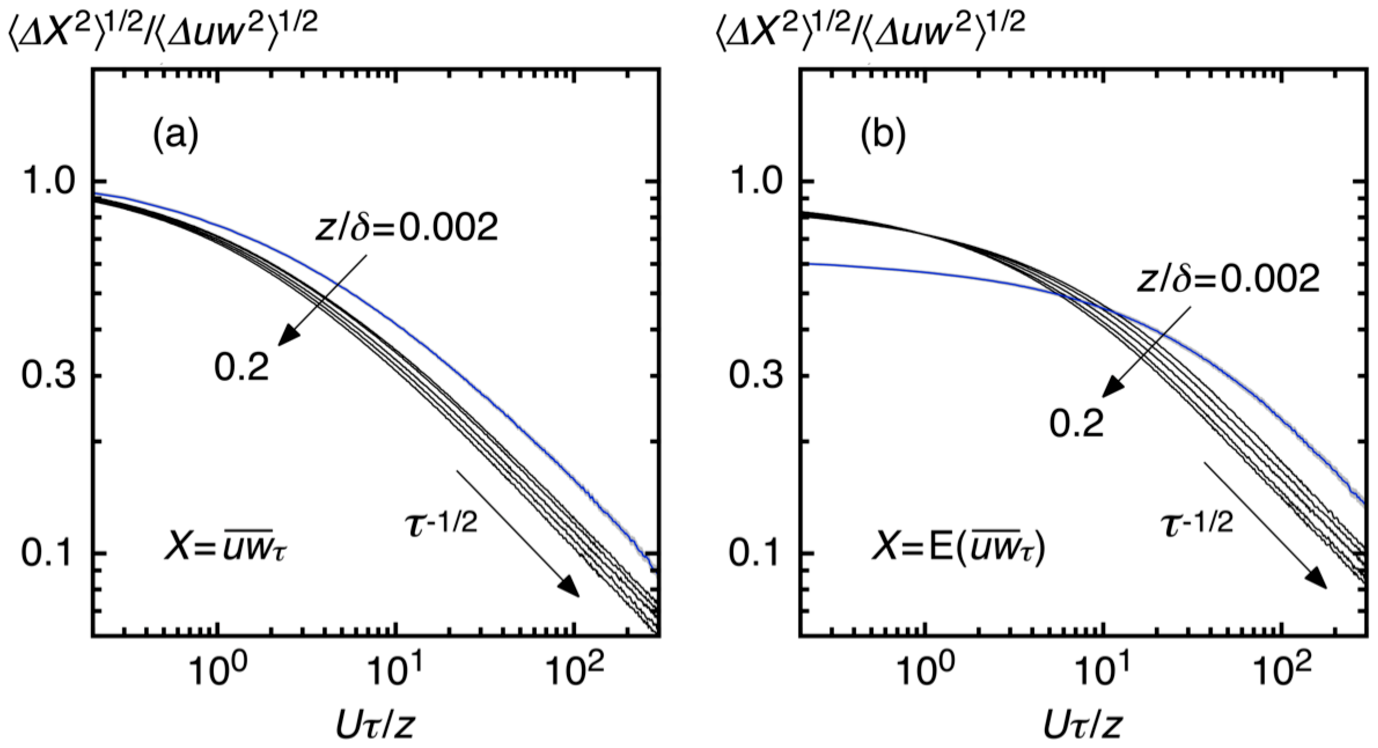}}
\caption{\label{f7} Same as in Fig.~\ref{f6} but for standard deviation $\langle {\mit\Delta}X^2 \rangle^{1/2}$ of ${\mit\Delta}X = X - \langle X \rangle = \overline{uw}_{\tau} - \langle \overline{uw}_{\tau} \rangle$ (a) or ${\rm E}(\overline{uw}_{\tau}) - \langle {\rm E}(\overline{uw}_{\tau}) \rangle$ (b) normalized by that of ${\mit\Delta} uw = uw - \langle uw \rangle$.}
\end{center}
\vspace{2mm}
\end{figure} 

Figure~\ref{f7} shows standard deviations of $\overline{uw}_{\tau}$ and ${\rm E}(\overline{uw}_{\tau})$ normalized by that of the instantaneous flux~$uw$.~These two are of similar magnitude. Even at the timescale~$\tau \simeq 20z/U$ for $z/\delta = 0.20$ to $60z/U$ for $z/\delta = 0.002$, the standard deviations of $\overline{uw}_{\tau}$ and ${\rm E}(\overline{uw}_{\tau}\!)$ are not small, $30$\% of that of $uw$. It is despite the fact that most fluctuations of the minor component $(u_{\perp}, w_{\perp})$ have been smoothed away and the statistics in Fig.~\ref{f6} have become almost constant. Some fluctuations of the major component $(u_{\shortparallel} , w_{\shortparallel})$ could have been smoothed also.

Regardless of the distance $z$, each standard deviation in Fig.~\ref{f7} obeys a power law $\tau^{-1/2}$ at the largest timescales $\tau$ (an arrow). This is because the two-time correlations of ${\mit\Delta}X = uw - \langle uw \rangle$ and $u^2 - \langle u^2 \rangle \varpropto {\rm E}(uw) - \langle {\rm E}(uw) \rangle$ are negligible there (Fig.~\ref{f4}). Toward a Gaussian distribution of ${\mit\Delta}\overline{X}_{\tau}$ in the course of $\tau \rightarrow +\infty$,\cite{my71} we have $\langle {\mit\Delta}\overline{X}_{\tau}^2 \rangle \varpropto \tau^{-1}$ and hence $\langle {\mit\Delta}\overline{X}_{\tau}^2 \rangle^{1/2} \varpropto \tau^{-1/2}$.

To conclude, the formulae of Eq.~(\ref{eq2}) hold anywhere in the constant-flux sublayer. The timescale $\tau$ for smoothing the minor component $(u_{\perp}, w_{\perp})$ needs to exceed a few tens of $z/U$, i.e., timescales for non-negligible two-time correlations of $uw$ and $u^2$. Between the estimate ${\rm E}(\overline{uw}_{\tau})$ and the actual value $\overline{uw}_{\tau}$, the correlation coefficient is $0.6$ to $0.7$. Its slope is unity. The standard deviation of ${\rm E}(\overline{uw}_{\tau})$ does not differ significantly from those of $uw$ and $\overline{uw}_{\tau}$.

\section{Discussion} \label{S5}

\begin{figure}[bp]
\begin{center}
\resizebox{7.9cm}{!}{\includegraphics*[2.7cm,14.6cm][18.2cm,22.6cm]{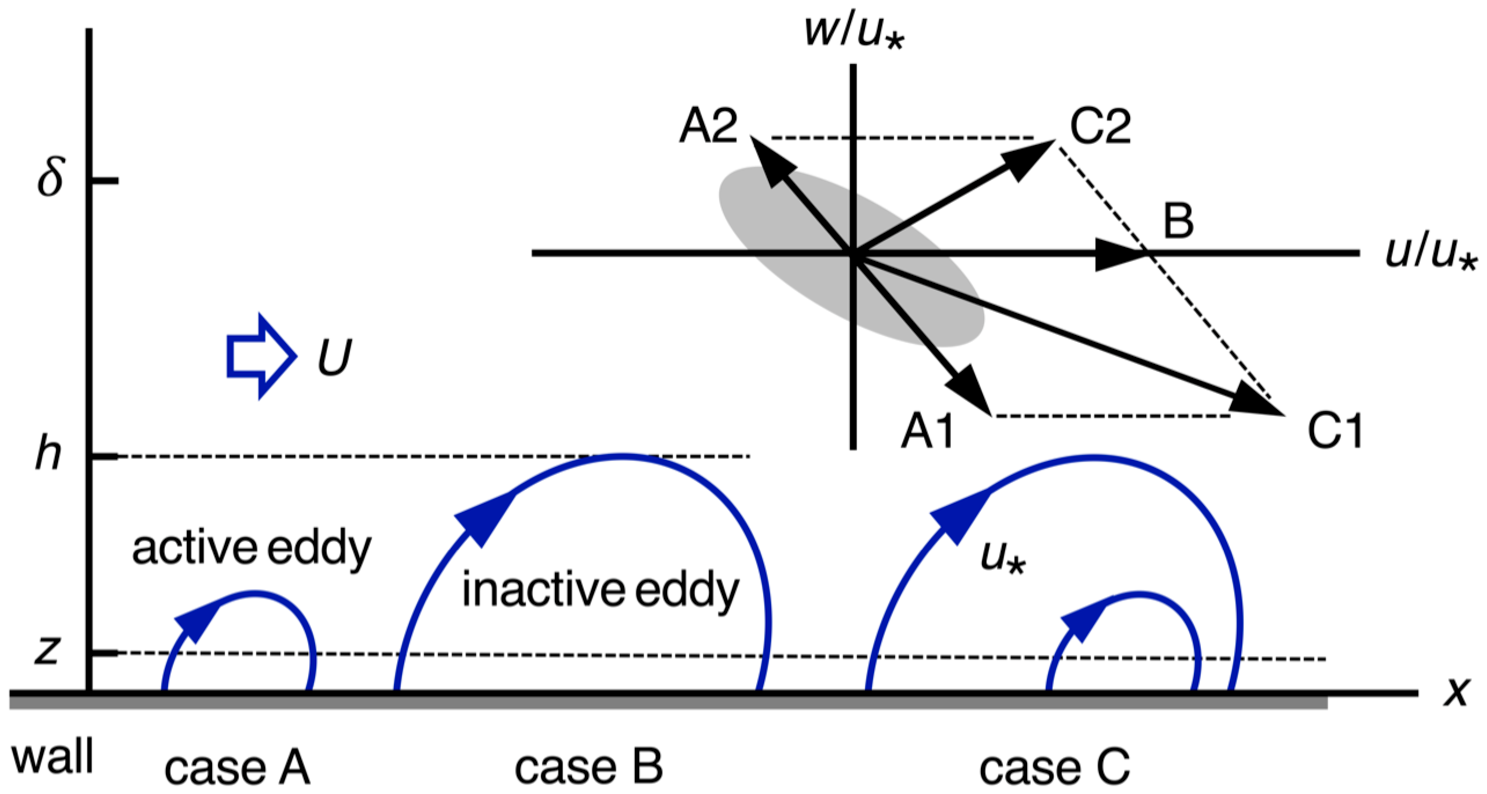}}
\caption{\label{f8} Schematic of the attached-eddy hypothesis. Case A: active eddy. Case B: inactive eddy. Case C: active eddy in an inactive eddy. The eddy size $h$ is for case B. We also show the velocity vectors $(u,w)$ at an observing distance $z$. That~for case C corresponds to a sum of those for cases A and B.}
\end{center}
\end{figure} 

\subsection{Interpretation on phenomenology of eddies} \label{S5a}

Having confirmed the formulae of Eq.~(\ref{eq2}), we interpret them on Townsend's attached-eddy hypothesis.\cite{t76, pc82, mm19} As illustrated in Fig.~\ref{f8}, this is a model of a random superposition of energy-containing eddies attached to the~wall. Such eddies do exist in wall turbulence,\cite{amt00, dn11b, mm19} e.g., packets of hairpin-shaped vortices.

These attached eddies have a common shape with the same characteristic velocity $u_{\ast}$. However, the size is not the same. The larger eddies are increasingly rare. Also, for the wall surface, we adopt a free-slip condition such that $u \nrightarrow 0$ and $v \nrightarrow 0$ but $w \rightarrow 0$ as $z \rightarrow 0$.

Since all the eddies are of the same shape and attached to the wall, $w$ is due only to those with wall-normal sizes $h$ comparable to the observing distance $z$, i.e., the active eddies\cite{t76} (Fig.~\ref{f8}, case A). They are responsible for the mean momentum flux $\langle uw \rangle = - u_{\ast}^2$. The larger eddies are inactive\cite{t76} and contribute to the wall-parallel velocities $u$ and $v$ alone (Fig.~\ref{f8}, case B).

The integration over all the attached eddies yields scaling laws of Eq.~(\ref{eq8a}) for $\langle u^2 \rangle$ and $\langle w^2 \rangle$ in the constant-flux sublayer.\cite{t76} While the constants $c_{uu}$ and $c_{ww}$ originate in those active eddies, the logarithmic term $d_{uu} \ln (\delta /z)$ originates in inactive eddies with sizes $h$ from the observing distance $z$ to the thickness $\delta$ of the turbulence region.

We calculate $\langle (uw - \langle uw \rangle )^2 \rangle = \langle u^2w^2 \rangle - \langle uw \rangle^2$ by using a cumulant $\langle X \rangle_c$.\cite{my71} From an identity $\langle X_1 X_2 X_3 X_4 \rangle_c = \langle X_1 X_2 X_3 X_4 \rangle - \langle X_1 X_2 \rangle \langle X_3 X_4 \rangle - \langle X_1 X_3 \rangle \langle  X_2 X_4 \rangle - \langle X_1 X_4 \rangle \langle X_2 X_3 \rangle$ for zero-mean random variables $X_1$, $X_2$, $X_3$, and $X_4$,
\begin{equation} \label{eq9}
\langle (uw - \langle uw \rangle )^2 \rangle = \langle u^2 w^2 \rangle_c + \langle uw \rangle^2 + \langle u^2 \rangle \langle w^2 \rangle.
\end{equation}
The cumulant at a position in the wall turbulence is just the sum of those for the individual attached eddies, e.g., $\langle (\sum_i Y_i)^n \rangle_c = \sum_i \langle Y_i^n \rangle_c$. For the case of $\langle u^2 w^2 \rangle_c$, which includes the wall-normal velocity $w$, it is enough to consider the active eddies alone. The result is a constant and written as $c_{uwuw}u_{\ast}^4$. Then, Eq.~(\ref{eq9}) leads via Eq.~(\ref{eq8a}) to Eq.~(\ref{eq8b}) as the scaling law of $\langle (uw - \langle uw \rangle )^2 \rangle$.

The constants $c_{uwuw}+1$ and $c_{uu}c_{ww}$ of Eq.~(\ref{eq8b}) originate respectively in single active eddies and their overlapping pairs. On the other hand, the logarithmic term $c_{ww} d_{uu} \ln (\delta/z)$ originates in overlapping pairs of active and inactive eddies (Fig.~\ref{f8}, case C), where an active eddy transfers the momentum $u$ of an inactive eddy. Nevertheless, such pairs cancel one another out and do not contribute to the mean momentum flux $\langle uw \rangle$.

When the distance $z$ is small, since inactive eddies are not rare, the fluctuations of $u$ and $uw$ are large. This is in agreement with our results in Figs.~\ref{f2} and \ref{f3}. Especially for the elliptical distribution of the data on the $u$--$w$ plane (Fig.~\ref{f2}), its inclination $K(z)$ decreases with an increasing contribution of those inactive eddies to the streamwise velocity $u$, i.e., $K(z) \varpropto u_{\ast}^2 / \langle u^2(z) \rangle = [ c_{uu} + d_{uu} \ln (\delta /z)]^{-1}$ in Eqs. (\ref{eq2c}) and (\ref{eq8a}).

\begin{table}[tbp]
\begingroup
\squeezetable
\caption{\label{t2} Energy partition between the major and the minor components and between the active and the inactive attached eddies in units of $u_{\ast}^2$. We also show $\pm 1 \sigma$ errors.}
\begin{ruledtabular}
\begin{tabular}{lccc}
Component & Eddy      & Laboratory        & Field              \\  
          &           & $z/\delta = 0.15$ & $z/\delta = 0.002$ \\  \hline
Major     & Active    & $3.00 \pm 0.08$   & $2.74 \pm 0.09$    \\    
          & Inactive  & $1.93 \pm 0.08$   & $8.09 \pm 0.18$    \\
Minor     & Active    & $1.03 \pm 0.02$   & $1.29 \pm 0.02$    \\
          & Inactive  & $0.31 \pm 0.01$   & $0.24 \pm 0.01$    \\ 
\end{tabular}
\end{ruledtabular}
\endgroup
\end{table}

To be quantitative, we analyze energies of the active and the inactive attached eddies in units of $u_{\ast}^2 = - \langle uw \rangle$, partitioned between the major component $(u_{\shortparallel} , w_{\shortparallel})$ and the minor component $(u_{\perp} , w_{\perp})$. From Eq.~(\ref{eq3}),
\begin{subequations}
\begin{align} \label{eq10}
&          \langle u_{\shortparallel}^2        \rangle +     \langle w_{\shortparallel}^2 \rangle =
 \frac{    \langle u^2 \rangle - 2K \langle uw \rangle + K^2 \langle w^2                  \rangle}{1+K^2} , \\
&          \langle u_{\perp}^2                 \rangle +     \langle w_{\perp}^2          \rangle =
 \frac{K^2 \langle u^2 \rangle + 2K \langle uw \rangle + \langle w^2 \rangle}{1+K^2} .
\end{align}
\end{subequations}
For the active eddies, we use our values of the constants $c_{uu}$ and $c_{ww}$ of Eq.~(\ref{eq8a}). The inactive eddies contribute only via the logarithmic term $d_{uu} \ln (\delta /z)$. We estimate its value from our $c_{uu}$ value along with $\langle u^2 \rangle/u_{\ast}^2$ measured at each distance $z$.

Table~\ref{t2} summarizes our results. Especially in the major component at the small distance of $z/\delta = 0.002$, the inactive attached eddies are predominant. The energies in the other categories are not different between the cases of $z/\delta = 0.002$ and $0.15$.

The streamwise size of an attached eddy is often assumed to be ${\cal O}(10)$ of its wall-normal size $h$.\cite{mm19, mmym20, hyz20} Because of $h \simeq z$ for the active attached eddies, the corresponding timescale $\tau$ is a few tens of $z/U$. At such a timescale $\tau$, most fluctuations of the minor component $(u_{\perp} , w_{\perp})$ have been smoothed away (Fig.~\ref{f6}). We attribute them to internal motions of those active eddies. Two-time correlations of $uw$ and $u^2$ are actually negligible there (Fig.~\ref{f4}).

The remaining fluctuations of the minor component $(u_{\perp} , w_{\perp})$ and also all the fluctuations of the major component $(u_{\shortparallel} , w_{\shortparallel})$ are due to net contributions of the active eddies as well as to the inactive eddies. With a decrease in $z/\delta$, the inactive eddies are increasingly predominant (Table~\ref{t2}). The smoothing timescale $\tau$ needs to be larger. When $z/\delta$ becomes too small, despite the correction of $K(z)$ in Eq.~(\ref{eq2c}), those inactive eddies tend to lower the accuracy of Eq.~(\ref{eq2a}) as in Fig.~\ref{eq6}(a).

\subsection{Application to numerical simulation} \label{S5b}

Here is a discussion on applying the formulae of Eq.~(\ref{eq2}) to a large-eddy simulation, i.e., a numerical simulation used widely for wall turbulence. It intends to resolve all energy-containing eddies, but such eddies are too small in the vicinity of the wall surface. Usually, the first off-wall grid points lie in the constant-flux sublayer. To estimate fluctuations of the momentum flux $uw$ there, a so-called wall model is necessary.\cite{d70, bmp05, kl12, lkbb16, ypm17, bp18, bl21, bogcn21}

At these grid points, the instantaneous streamwise velocity $U+u$ is still available. Its value is often used to estimate $uw$, by replacing $U$ with $U+u$ and $\langle uw \rangle$ with $uw$ in the law of the wall of Eq.~(\ref{eq1b}).\cite{d70}

However, such a simulation does not reproduce the profile of the mean velocity $U(z)$. This discrepancy depends on numerical details,\cite{kl12, lkbb16, bp18, bl21, bogcn21} e.g., subgrid-scale modeling, but the main reason is that Eq.~(\ref{eq1b}) is not to relate $U+u$ with $uw$ but $U$ with $\langle uw \rangle$.\cite{bmp05, ypm17, bp18, im21} We could have $U+u$ and $uw$ close to their averages $U$ and $\langle uw \rangle$, by~smoothing $U+u$ and $uw$ in space\cite{bmp05, ypm17} or in time\cite{ypm17} or by focusing~on $uw$ and $U+u$ further away from the wall.\cite{kl12}

The formulae of Eq.~(\ref{eq2}) are about fluctuations of $uw$ themselves and hence serve as a more realistic model. For example, although we only obtain $uw < 0$ from the law of the wall of Eq.~(\ref{eq1b}), our law of Eq.~(\ref{eq2a}) could also induce $uw > 0$ to enhance the turbulence (Figs.~\ref{f2} and \ref{f5}). The wall-normal distance $z$ of those grid points differs among the cases, for which the factor $K(z)$ is always corrected via Eq.~(\ref{eq2c}).

For the smoothing timescale $\tau$, a few tens of $z/U$ is appropriate. Here, compromise is necessary between the estimation accuracy and the fluctuation magnitude. The former is higher at a larger $\tau$ (Fig.~\ref{f6}), while the latter is larger at a smaller $\tau$ (Fig.~\ref{f7}).

Even in a non-stationary case such as the atmospheric boundary layer, Eq.~(\ref{eq2}) is valid so long as the mean flow direction and the mean flow velocity $U$, among others, are well-defined over some averaging timescale, e.g., $15$\,min for our field data in Sec.~\ref{S3b}. It needs to by far exceed a few tens of $z/U$, i.e., the smoothing timescale $\tau$.

Though we have studied boundary layers alone, Eq.~(\ref{eq2}) also applies to other types of wall turbulence, e.g., pipe flows, since the constant-flux sublayer remains essentially the same. The estimation accuracy and the appropriate timescale $\tau$ might yet differ. In fact, while the parameter $c_{uu}$ of Eq.~(\ref{eq8a}) is $2.0$--$2.5$ in boundary layers,\cite{mmhs13, smhffhs18, mmym20} it is $1.4$--$1.9$ in pipes.\cite{hvbs12, mmhs13, ofsbta17, ofwkt22} For the latter, we need to use the pipe radius as the thickness $\delta$ of the turbulence region.

Finally, we discuss the wall-normal flux of the spanwise momentum $vw$. The average $\langle vw \rangle$ is almost always negligible, but fluctuations of $vw$ are important to simulating the instantaneous spanwise velocity $v$. Since the major component $w_{\shortparallel} = -K u_{\shortparallel}$ of Eq.~(\ref{eq3}) transfers that momentum as $v w_{\shortparallel} = -K u_{\shortparallel} v$, some calculations like Eqs.~(\ref{eq5})--(\ref{eq7}) yield a law analogous to Eq.~(\ref{eq2a}),
\begin{equation} \label{eq11}
\overline{vw}_{\tau}(z,t) - \langle vw \rangle = -K(z) \left[ \overline{uv}_{\tau}(z,t) - \langle uv \rangle \right] .
\end{equation}
Here $\langle vw \rangle$ and $\langle uv \rangle$ are not necessarily equal to $0$ if the flow direction is not exactly stationary.

\begin{figure}[tbp]
\begin{center}
\resizebox{8.8cm}{!}{\includegraphics*[3.0cm,18.7cm][17.3cm,26.6cm]{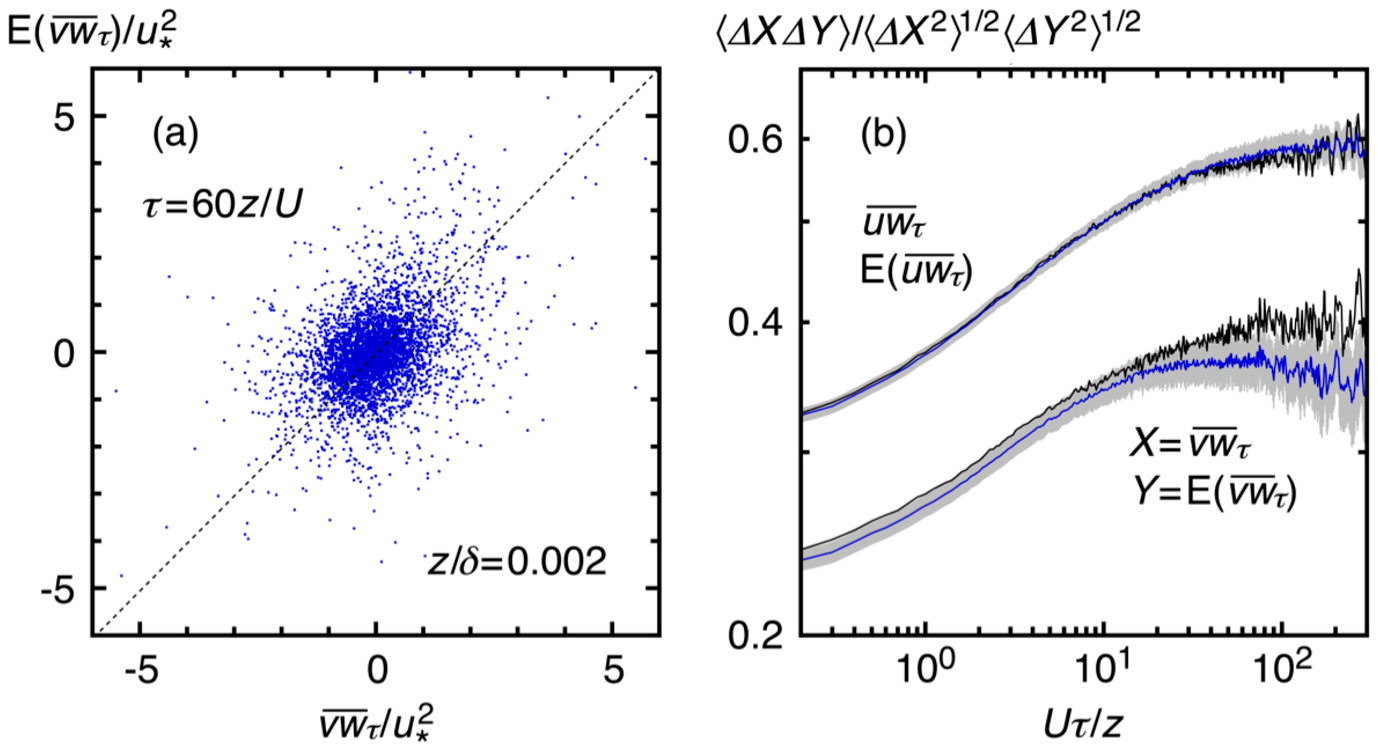}}
\caption{\label{f9} Scatter plot of ${\rm E}(\overline{vw}_{\tau})/u_{\ast}^2$ and $\overline{vw}_{\tau}/u_{\ast}^2$ for $\tau = 60z/U$ (a) and correlation coefficient $\langle {\mit\Delta}X{\mit\Delta}Y \rangle / \langle {\mit\Delta}X^2 \rangle^{1/2} \langle {\mit\Delta}Y^2 \rangle^{1/2}$ with ${\mit\Delta}X = X - \langle X \rangle$ and ${\mit\Delta}Y = Y - \langle Y \rangle$ between $X = \overline{vw}_{\tau}$ or $\overline{uw}_{\tau}$ and $Y = {\rm E}(\overline{vw}_{\tau})$ or ${\rm E}(\overline{uw}_{\tau})$ against $U \tau/z$ (b) at $z/\delta = 0.002$. Here $Y = {\rm E}(X)$ is an estimate of $X$ via Eq.~(\ref{eq2}) or (\ref{eq11}). Gray areas denote $\pm 1\sigma$ errors. The curves without the errors are from a more limited criterion for the data selection.}
\end{center}
\end{figure}

Figure~\ref{f9} shows the results for our field data, where $u$, $v$, and $w$ are all available (Sec.~\ref{S3b}). Upon a scatter plot of the estimate ${\rm E}(\overline{vw}_{\tau})$ against the actual value $\overline{vw}_{\tau}$ in Fig.~\ref{f9}(a), the data points lie along a dotted line of $\overline{vw}_{\tau} = {\rm E}(\overline{vw}_{\tau})$. The correlation coefficient in Fig.~\ref{f9}(b) is $0.3$ to $0.4$ at $\tau \simeq 60z/U$. It decreases as $\tau$ increases still more, owing to large-timescale variations of the wind direction. If its standard deviation is limited to $< 16^{\circ}$ instead of $< 20^{\circ}$ of the original data,\cite{mmh19} we have $43$ data blocks. They do not yield that decrease (a curve without errors). Between $\overline{uw}_{\tau}$ and ${\rm E}(\overline{uw}_{\tau})$, the correlation coefficient in Fig.~\ref{f9}(b) remains the same.

Being thus responsible for at least some fluctuations of the flux $vw$, Eq.~(\ref{eq11}) is of use in a large-eddy simulation. As for the non-stationary case, $\langle vw \rangle$ and $\langle uv \rangle$ are difficult to estimate. Nevertheless, since they are small, if any, an assumption of $\langle vw \rangle = \langle uv \rangle = 0$ would be enough.

\section{Conclusion and Future Issue} \label{S6}

For the constant-flux sublayer of wall turbulence, we have formulated Eq.~(\ref{eq2}) to relate temporal fluctuations of the momentum flux $uw$ to those of the streamwise velocity $U+u$. The flux $uw$ is dominated by the major component $w_{\shortparallel} = -K u_{\shortparallel}$ on the $u$--$w$ plane (Fig.~\ref{f2}). Here $K$ is a function of the wall-normal distance $z$ as in Eq.~(\ref{eq2c}). Via smoothing in Eq.~(\ref{eq2b}) to single out that component, we have derived Eq.~(\ref{eq2a}). It is consistent with our data of boundary layers across a wide range of $z/\delta = 0.002$ to $0.2$ (Figs.~\ref{f5}--\ref{f7}). The smoothing timescale $\tau$ needs to exceed a few tens of $z/U$, i.e., timescales for non-negligible two-time correlations of $uw$ and $u^2$ (Fig.~\ref{f4}). Also for the flux $vw$, we have formulated Eq.~(\ref{eq11}) and confirmed it on our data (Fig.~\ref{f9}). These Eqs.~(\ref{eq2}) and (\ref{eq11}) apply readily to wall modeling of a large-eddy simulation.

The formulae of Eq.~(\ref{eq2}) originate in our previous study of Ref.~\onlinecite{mi22}. Only with laboratory data at $z/\delta = 0.1$ to $0.2$, we assumed that $K$ in Eq.~(\ref{eq2a}) was identical to the von K\'arm\'an constant $\kappa$. In fact, $K$ varies around $\kappa$. By also using field data at $z/\delta = 0.002$ of Ref.~\onlinecite{mmh19} (Table~\ref{t1} and Fig. \ref{f3}), we have reconsidered the functional form of $K$. The formulae of Eq.~(\ref{eq2}) now hold anywhere in the constant-flux sublayer and agree with a phenomenology of active and inactive attached eddies\cite{t76} (Table~\ref{t2}).

The following is a remark on a future issue. We have focussed on thermally neutral cases. If the wall is horizontal and chilled or heated against the overlying flow, the turbulence is stable or unstable,\cite{my71, f06, m14, ypa20} as is usual in the atmospheric boundary layer. Although the constant-flux sublayer remains, we need to extend its law of Eq.~(\ref{eq1a}). If $X$ is the momentum, temperature, or passive scalar concentration, the wall-normal flux is $w(X - \langle X \rangle) = w {\mit\Delta}X$. The average $\langle w {\mit\Delta}X \rangle$ is constant in that sublayer. Then, according to Monin and Obukhov,\cite{my71,f06}
\begin{equation} \label{eq12}
\frac{z}{{\mit\Delta}X_{\ast}} \frac{d \langle X \rangle} {dz} = \phi_X \! \left( \frac{z}{L_{\rm O}} \right) 
\  \mbox{with} \ 
{\mit\Delta}X_{\ast} = \frac{| \langle w {\mit\Delta}X \rangle |}{u_{\ast}} .
\end{equation}
Here $\phi_X$ is a non-dimensional function. The Obukhov length $L_{\rm O}$ is such that $L_{\rm O} > 0$ in stable cases, $L_{\rm O} < 0$ in unstable cases, and $L_{\rm O} \rightarrow \pm\infty$ in the neutral limit. Except under too stable conditions,\cite{m14} Eq.~(\ref{eq12}) is always reliable.

By using ${\mit\Delta}X_{\ast}$ and $\phi_X$ of Eq.~(\ref{eq12}) instead of $u_{\ast}$ and $1/\kappa$ of Eq.~(\ref{eq1a}), we extend our formulae of Eq.~(\ref{eq2}). The major component  on the ${\mit\Delta}X$--$w$ plane is $w_{\shortparallel} = -K_X {\mit\Delta}X_{\shortparallel}$. From calculations like Eqs.~(\ref{eq4})--(\ref{eq7}),
\begin{subequations} \label{eq13}
\begin{equation} \label{eq13a}
\overline{w{\mit\Delta}X}_{\tau}(z,t) - \langle w{\mit\Delta}X \rangle = -K_X(z) \! \left[ \overline{{\mit\Delta}X^2}_{\tau}(z,t) - \langle {\mit\Delta}X^2(z) \rangle \right] \!.
\end{equation}
The proportionality factor $K_X$ depends on the distance $z$ as
\begin{equation} \label{eq13b}
K_X(z) = \frac{1}
              {\phi_X(z_{\rm m}/L_{\rm O})}
         \frac{\langle w{\mit\Delta}X \rangle / \langle {\mit\Delta}X^2(z)         \rangle} 
              {\langle w{\mit\Delta}X \rangle / \langle {\mit\Delta}X^2(z_{\rm m}) \rangle}.
\end{equation}
\end{subequations}
Here $\tau$ is to be determined with $U$, $z$, and also $L_{\rm O}$. The empirical values of $\phi_X(z_{\rm m}/L_{\rm O})$ and $\langle w{\mit\Delta}X \rangle / \langle {\mit\Delta}X^2(z_{\rm m}) \rangle$ are available in the literature.\cite{my71, f06, ypa20} It would be of interest to study Eq.~(\ref{eq13}) with laboratory or field data toward a future application to a large-eddy simulation of thermally stratified wall turbulence.

\section*{Appendix \ \ Correlation Functions} \label{SA}

For wall turbulence that is homogeneous in the $x$ direction, the attached-eddy hypothesis yields scaling laws of two-point correlations at any distance $z$ in the constant-flux sublayer. With no additional assumption, laws analogous to Eq.~(\ref{eq8a}) are derived as\cite{mouri17, mmym20}
\begin{subequations} \label{eq14}
\begin{align} 
\frac{\langle u(x+r) u(x) \rangle}{u_{\ast}^2} &= C_{uu} \! \left( \frac{r}{z} \right) + D_{uu} \! \left( \frac{r}{z} \right) \ln \left( \frac{\delta}{z} \right), \\
\frac{\langle w(x+r) w(x) \rangle}{u_{\ast}^2} &= C_{ww} \! \left( \frac{r}{z} \right), \\
\frac{\langle u(x+r) w(x) \rangle}{u_{\ast}^2} &= C_{uw} \! \left( \frac{r}{z} \right).
\end{align}
\end{subequations}
We have used $\langle X \rangle$ for $\langle X(z) \rangle$. The functions $C_{uu}$, $C_{ww}$, $C_{uw}$, and $D_{uu}$ have the values $C_{uu}(0) = c_{uu}$, $C_{ww}(0) = c_{ww}$, $C_{uw}(0) = -1$, and $D_{uu}(0) = d_{uu}$. Then, as in our derivation of Eq.~(\ref{eq8b}) via Eq.~(\ref{eq9}),
\begin{subequations} \label{eq15}
\begin{align} \label{eq15a}
&\frac{ \langle uw(x+r)uw(x) \rangle - \langle uw \rangle^2}{u_{\ast}^4}        \nonumber \\
&\ = \frac{ \langle uw(x+r)uw(x) \rangle_c }{u_{\ast}^4} 
   + \frac{\langle u(x+r)w(x) \rangle \langle u(x)w(x+r) \rangle }{u_{\ast}^4}  \nonumber \\
&\ + \frac{\langle u(x+r) u(x) \rangle \langle w(x+r) w(x) \rangle}{u_{\ast}^4}, 
\end{align}
with a cumulant
\begin{equation}
\frac{\langle uw(x+r)uw(x) \rangle_c}{u_{\ast}^4} = C_{uwuw} \! \left( \frac{r}{z} \right).
\end{equation}
\end{subequations}
The individual terms of Eq.~(\ref{eq15}) correspond to those of Eqs.~(\ref{eq8b}) and (\ref{eq9}), e.g., $C_{uwuw}(0) = c_{uwuw}$. Likewise,\cite{mmym20}
\begin{subequations} \label{eq16}
\begin{align} \label{eq16a}
&     \frac{  \langle u^2(x+r)u^2(x) \rangle - \langle u^2 \rangle^2}{u_{\ast}^4} \nonumber \\
&\ =  \frac{  \langle u^2(x+r)u^2(x) \rangle_c}{u_{\ast}^4}
    + \frac{2 \langle u(x+r)  u(x)   \rangle^2}{u_{\ast}^4},
\end{align}
with another cumulant
\begin{equation} \label{eq16b}
\frac{\langle u^2(x+r)u^2(x) \rangle_c}{u_{\ast}^4} = C_{u^2u^2} \! \left( \frac{r}{z} \right) + D_{u^2u^2} \! \left( \frac{r}{z} \right) \ln \left( \frac{\delta}{z} \right).
\end{equation}
\end{subequations}
Taylor's hypothesis\cite{my75} $r = -U\tau$ converts the two-point correlations of Eqs.~(\ref{eq14})--(\ref{eq16}) into two-time correlations. These are functions of $U\tau/z$ and $\ln (\delta/z)$. Since $\ln (\delta/z)$ is usually ${\cal O}(1)$ in the constant-flux sublayer, it does not significantly affect the timescale $\tau$ for the disappearance of each of the correlations as observed in Fig.~\ref{f4}.

\begin{acknowledgments} 
This study was supported in part by KAKENHI Grant No.~19K03967. We are grateful to anonymous referees for their valuable comments on the manuscript.
\end{acknowledgments}

\section*{Author Declarations}

\noindent {\bf Conflict of Interest:} The authors have no conflicts to disclose.  

\noindent {\bf Data Availability:} Data sharing does not apply to this study, which obtained no new data.

\end{document}